\documentclass[11pt,dvips]{article}

\usepackage{epsfig,times} 
\usepackage{picinpar}
\usepackage{wrapfig}
\usepackage{floatflt}

\makeatletter
\renewcommand{\section}{\@startsection{section}{1}{0in}
	{0.4\baselineskip}{0.1\baselineskip}{\Large\bf}}
\renewcommand{\subsection}{\@startsection{subsection}{2}{0in}
	{0.25\baselineskip}{-\baselineskip}{\large\bf}}
\renewcommand{\subsubsection}{\@startsection{subsubsection}{3}{0in}
	{0.1\baselineskip}{-\baselineskip}{\normalsize\bf}}
\makeatother
\newcommand{\icrc}{$26^{\rm th}$ ICRC\ (Salt Lake City, 1999)}
\def\gray{$\gamma$-ray\ }
\def\grays{$\gamma$-rays\ }
\def\Berat{$^{10}$Be/$\,^9$Be}
\newcommand{\sigv}{\langle\sigma v\rangle}
\def\fwa{8cm}
\def\fwb{10cm}

\newcommand{\apj}{ApJ\ }
\newcommand{\astroph}[1]{{astro-ph/#1}}
\begin{document}

%
\makeatletter\newcommand{\ps@icrc}{
\renewcommand{\@oddhead}{Proc.~\icrc, \slshape{HE 5.1.15}\hfil}}
\makeatother\thispagestyle{icrc}
%
%

\begin{center}
{\LARGE \bf 
Galactic propagation of positrons from particle dark-matter annihilation}
\end{center}

\begin{center}
{\bf Igor V.~Moskalenko$^{1,2}$ and Andrew W.~Strong$^{1}$}\\
{\it $^{1}$MPI f\"ur extraterrestrische Physik, D--85740 Garching, Germany\\
$^{2}$Institute for Nuclear Physics, Moscow State University, 119 899 Moscow, 
   Russia}
\end{center}

\begin{center}
{\large \bf Abstract\\}
\end{center}
\vspace{-0.5ex}
We have made a calculation of the propagation of positrons from
dark-matter particle annihilation in the Galactic halo for different
models of the dark matter halo distribution using our 3D code. We show
that the Green's functions are not very sensitive to the dark matter
distribution for the same local dark matter energy density.  We compare
our predictions with computed cosmic ray positron spectra
(``background'') for the ``conventional'' cosmic-ray (CR) nucleon
spectrum which matches the local measurements, and a modified spectrum
which respects the limits imposed by measurements of diffuse Galactic
$\gamma$-rays, antiprotons, and positrons. We conclude that significant
detection of a dark matter signal requires favourable conditions and
precise measurements unless the dark matter is clumpy which would
produce a stronger signal.  Although our conclusion qualitatively
agrees with that of previous authors, it is based on a more realistic
model of particle propagation and thus reduces the scope for future
speculations.  Reliable background evaluation requires new accurate
positron measurements and further developments in modelling production
and propagation of cosmic ray species in the Galaxy.

\vspace{1ex}

%
%
\section{Introduction:} \label{intro}

Investigations of galaxy rotation, big-bang nucleosynthesis, and
large-scale structure formation imply that a significant amount of the
mass of the universe consists of non-luminous dark matter (Trimble
1989).  Among the favored particle dark matter candidates are so-called
weakly interacting massive particles (WIMPs), whose existence follows
from supersymmetric models (see Jungman, Kamionkowski, \& Griest 1996
for a review).  A pair of stable WIMPs can annihilate into known
particles and antiparticles and it may be possible to detect WIMPs in
the Galactic halo by the products of their annihilations.  Though the
microphysics is quite well understood and many groups make
sophisticated calculations of the spectra of annihilation products for
numerous WIMP candidates which include many decay chains (e.g., Baltz
\& Edsj\"o 1998),  there are still  uncertainties in the macrophysics
which could change the estimated fluxes of WIMP annihilation products
by 1--2 orders of magnitude, making  predictions for their detection
difficult.  The most promising is perhaps the positron signal since it
can appear at high energies where the solar modulation is negligible,
but its strength depends on many details of propagation in the Galaxy.
The ``leaky box'' model is often used (e.g., Kamionkowski \& Turner
1991), a simplified approach which may not be applicable in the case of
positrons.  On the other hand, progress in CR positron measurements is
anticipated since several missions operating or under construction are
capable of measuring positron fluxes up to 100 GeV
(e.g.\ experiments gas-RICH/CAPRICE: Barbiellini et al.\  1997, and PAMELA:
Adriani et al.\ 1997).  Therefore, more accurate
calculation of the positron propagation is desirable.

We have developed a numerical method and corresponding computer code
(GALPROP) for the calculation of Galactic CR propagation in 3D (for an
overview of our approach and results see Strong \& Moskalenko 1999, and
also papers OG~2.4.03, OG~3.2.18 in proceedings of this conference).
Briefly, the idea is to develop a model which simultaneously reproduces
observational data of many kinds related to cosmic-ray origin and
propagation: directly via measurements of nuclei, antiprotons,
electrons, and positrons, indirectly via \grays and synchrotron
radiation.  Here we use our model for calculation of positron
propagation in different models of the dark matter halo distribution
(Moskalenko \& Strong 1999).  To be specific we will discuss neutralino
dark matter, although our results can be easily adopted for any other
particle dark matter candidate.

\section{Green's functions:} \label{green}

The positron flux at the solar position is given by
\begin{equation}
\label{eq.3}
\frac{dF}{dE}= \int d\epsilon\, G(E,\epsilon) \sum_i B_i f_i(\epsilon)
\quad [{\rm cm}^{-2} {\rm\ s}^{-1} {\rm\ sr}^{-1} {\rm\ GeV}^{-1}],
\end{equation}
where $f(\epsilon)$ is the source function which describes the spectrum
of positrons from neutralino annihilation, $G(E,\epsilon)$ is the
Green's function for positron propagation in the Galaxy, and $B_i$ is
the branching ratio into a given final state $i$.  The Green's function
thus includes all details of the dark matter mass distribution,
neutralino annihilation cross section, and Galactic structure
(diffusion coefficient, spatially and energy dependent energy losses
etc.). We can write it in the form:
\begin{equation}
\label{eq.4}
G(E,\epsilon)= \sigv \frac{\rho_0^2}{m_\chi^2} \,
g(E,\epsilon)
\quad [{\rm cm}^{-2} {\rm\ s}^{-1} {\rm\ sr}^{-1} {\rm\ GeV}^{-1}],
\end{equation}
where $\sigv$ is the thermally averaged
annihilation cross section, $\rho_0$ is the local dark matter mass
density, $m_\chi$ is the neutralino mass, and we have introduced a function
$g(E,\epsilon)$ which describes the positron propagation for a
\begin{wrapfigure}[32]{r}{95mm}
\psfig{file=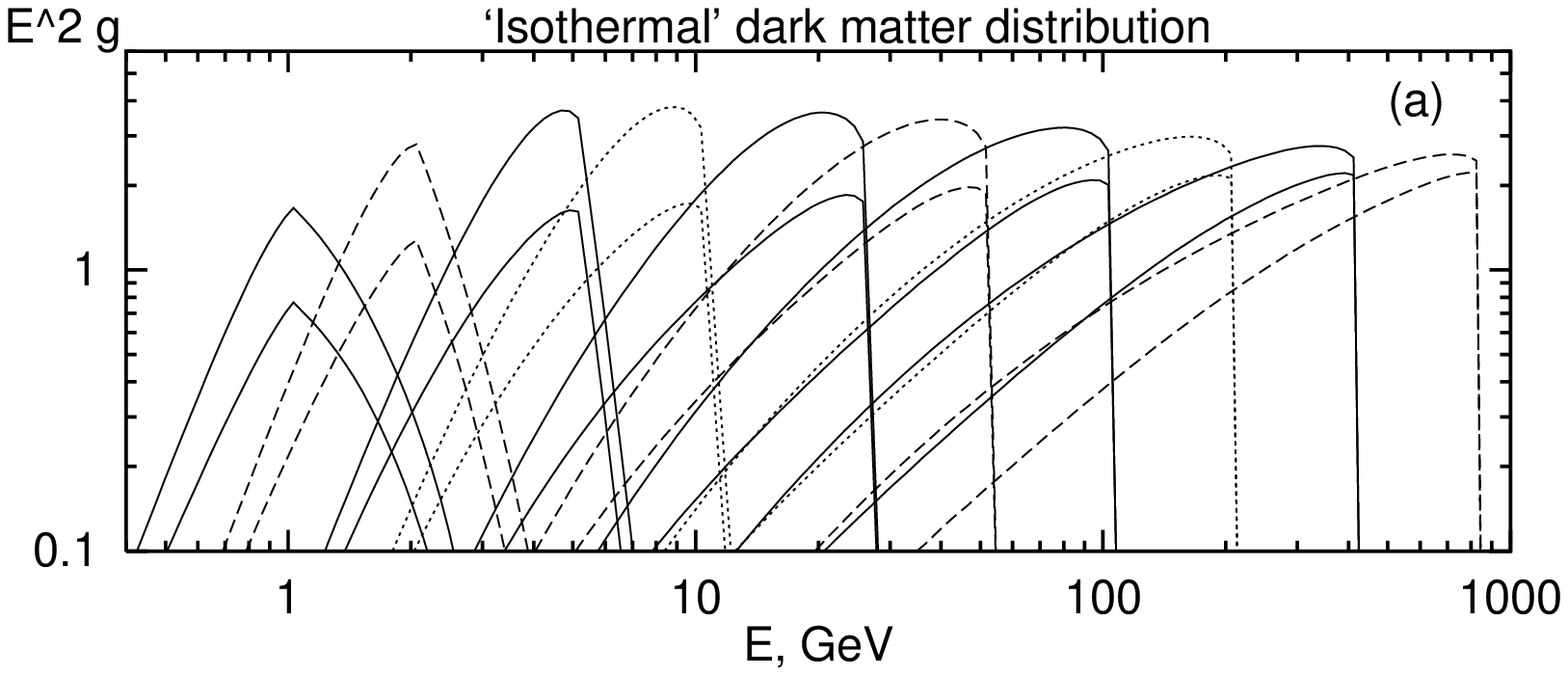,width=\fwb,clip=}\\
\psfig{file=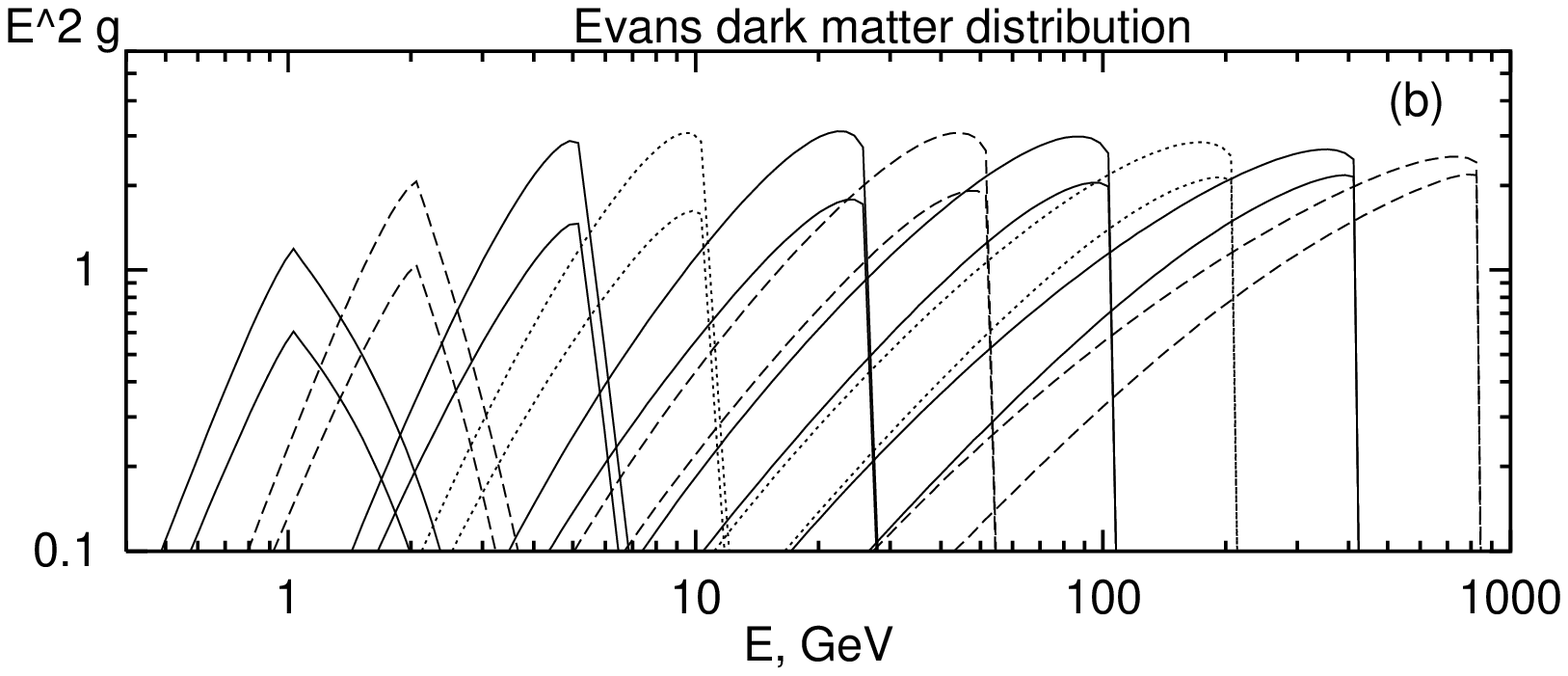,width=\fwb,clip=}\\
\psfig{file=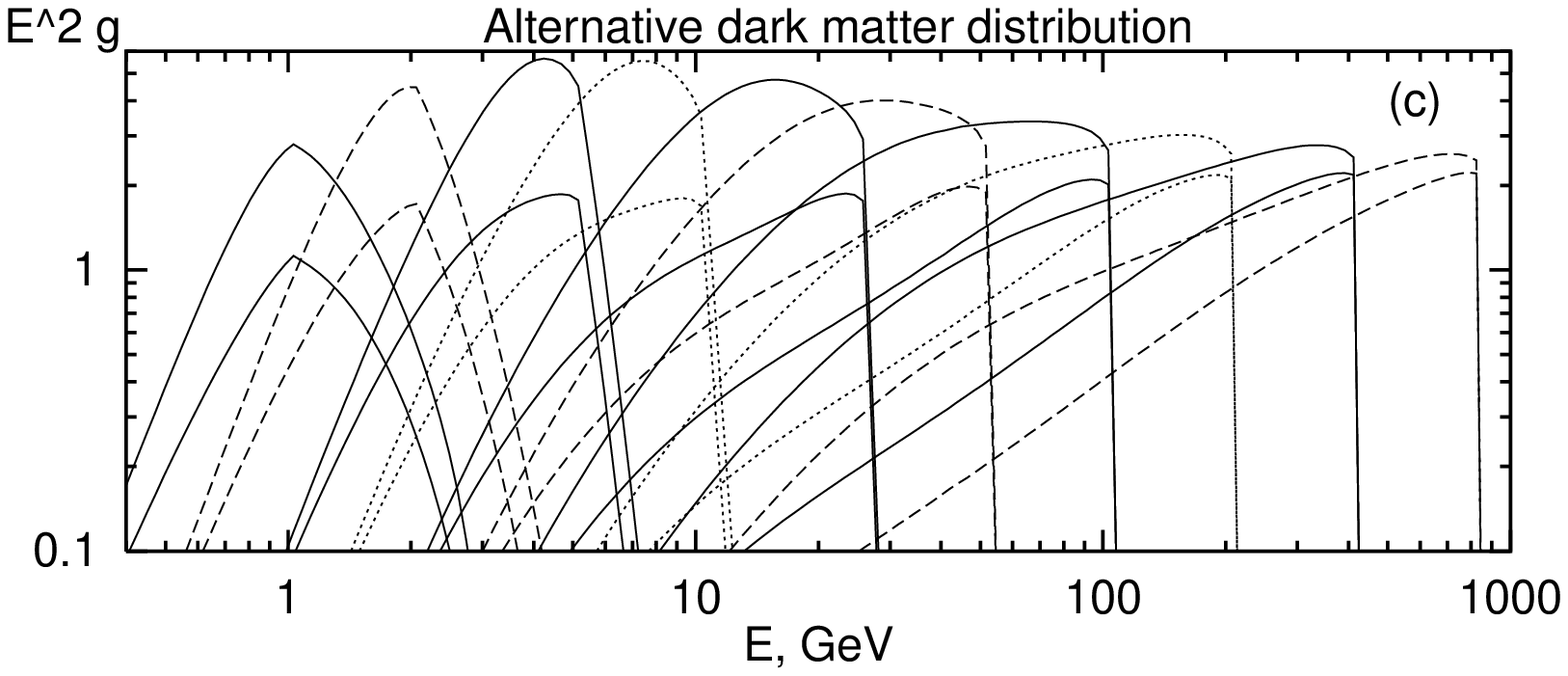,width=\fwb,clip=}
\caption[HE_5.1.15_f1a.ps,HE_5.1.15_f1b.ps,HE_5.1.15_f1c.ps]{ 
Calculated $g$-functions for different models of the dark matter
distribution: (a) ``isothermal'', (b) Evans, (c) alternative.  Upper
curves $z_h = 10$ kpc, lower curves $z_h = 4$ kpc, $\epsilon = 1.03$,
$2.06$, $5.15$, $10.3$, $25.8$, $51.5$, $103.0$, $206.1$, $412.1$,
$824.3$ GeV.  The units of the abscissa are $10^{25}$ GeV cm sr$^{-1}$.
\label{fig1} }
\end{wrapfigure}
given dark matter mass density distribution in the halo.

Following Kamionkowski and Kinkhabwala (1998) we consider three
different dark matter mass density profiles which match the Galactic
rotation curve. The canonical ``isothermal'' sphere profile, the
spherical Evans model, and an alternative model.
For each given model we calculate the function $g(E,\epsilon)$ defined
in Eq.~(\ref{eq.4}), which gives the positron flux at the solar
position corresponding to the positron source function in the form of a
Dirac $\delta$-function in energy.  The positron propagation is
calculated in a model which was tuned to match many available
astrophysical data (Strong \& Moskalenko 1998, Strong, Moskalenko, \&
Reimer 1999). Since the halo size in the range $z_h=4-10$ kpc is
favored by our analyses of B/C and \Berat\ ratios and diffuse Galactic
\gray emission, we consider two cases $z_h=4$ and
$10$ kpc which  provide us with an idea of the possible limits. The
preferred neutralino mass range following from accelerator and
astrophysical constraints is $50$ GeV $< m_\chi
<600 $ GeV (Ellis 1998), and we consider positron energies
$\epsilon \le 824$ GeV which cover this range.

Fig.~\ref{fig1} shows our calculated $g$-functions for different models
of the dark matter distribution: ``isothermal'', Evans, and
alternative. The curves are shown for two halo sizes $z_h = 4$ and $10$
kpc and several energies $\epsilon = 1.03$, $2.06$, $5.15$, $10.3$,
$25.8$, $51.5$, $103.0$, $206.1$, $412.1$, $824.3$ GeV.  At high
energies, increasing positron energy losses due to the inverse Compton
scattering compete with the increasing  diffusion coefficient, while at
low energies increasing energy losses due to the Coulomb scattering and
ionization (Strong \& Moskalenko 1998) compete with energy gain due to
reacceleration.  The first effect leads to a smaller sensivity to the
halo size at high energies.  The second one becomes visible below $\sim
5$ GeV and is responsible for the appearance of accelerated particles
with $E>\epsilon$.

It is interesting to note that for a given initial positron energy all
three dark matter distributions provide very similar values for the
maximum of the $g$-function (on the $E^2 g(E,\epsilon)$ scale), while
their low-energy tails are different.  This is a natural consequence of
the large positron energy losses.  Positrons contributing to the
maximum of the $g$-function  originate in the solar neighbourhood,
where all models give the same dark matter mass density.  The central
mass density in these models is very different, and therefore the shape
of the tail is also different since it is produced by positrons
originating in  distant regions.  As compared to the isothermal model,
the Evans model produces sharper tails, while the alternative model
gives more positrons in the low-energy tail.  At intermediate energies
($\sim10$ GeV) where the energy losses are minimal, the difference
between $z_h=4$ and 10 kpc is maximal.  Also at these energies
positrons from  dark matter particle annihilations in the Galactic
center can contribute to the predicted flux.  This is clearly seen in
the case of the alternative model with its very large central mass
density (Fig.~\ref{fig1}c, $z_h=10$ kpc).

\section{Positron fluxes:}  \label{flux}

An important issue in the interpretation of the positron measurements is
the evaluation of the ``background'', pos\-it\-rons arising from CR particle
interactions with interstellar matter.  Though the parameters of the
propagation and the Galactic halo size can be fixed in a
self-consistent way using CR isotope ratios, the ambient CR proton
spectrum on the Galactic scale remains quite uncertain.
The only possibility to trace the spectrum of nucleons on a large scale
is to observe secondary products such as diffuse $\gamma$-rays,
positrons, and antiprotons. 

\begin{figwindow}[3,r,
{\mbox{\psfig{file=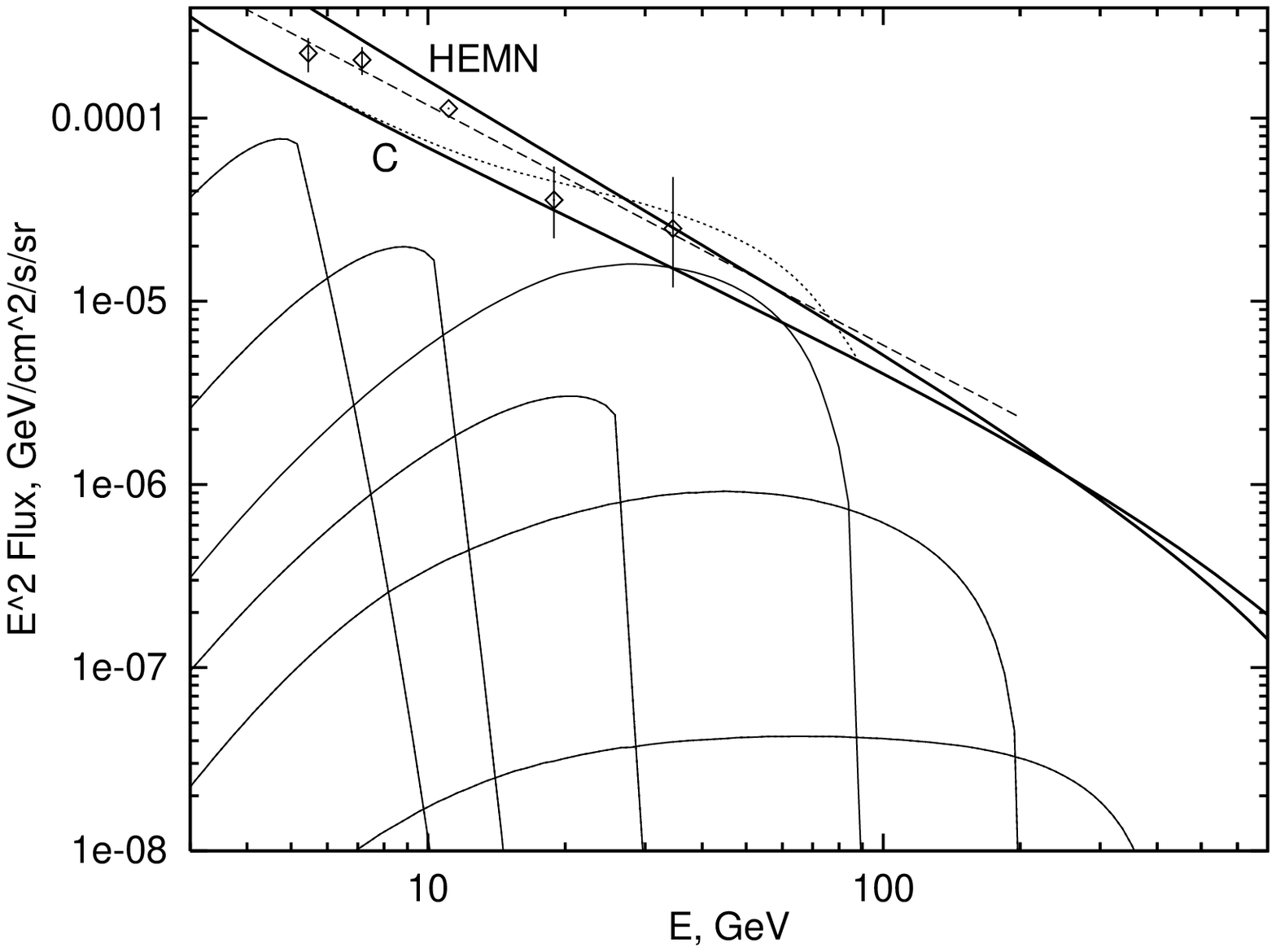,width=\fwa,clip=}}},%
{
Our predictions for two CR positron ``background'' models (C and HEMN:
heavy solid lines), and positron signals from neutralino annihilation
for $m_\chi=5.15$, $10.3$, $25.8$, $103.0$, $206.1$, $412.1$ (thin
solid lines):  (a) $z_h=4$ kpc, (b) $z_h=10$ kpc.  In the case of
$m_\chi=103.0$ GeV, the signal plus background (model C) is shown by
the dotted line.  Data and the best fit to the data (dashes) are from
Barwick et al.\ (1998, HEAT collaboration).
\label{fig2}}
]
In order to show the effect of varying of the ambient proton spectrum,
we compare our results with two models for the CR positron ``background''.
These are a ``conventional'' model (model C) which reproduces the local
directly measured proton and Helium spectra above 10 GeV (where solar
modulation is small), and a model with modified nucleon spectrum (model
HEMN), which is flatter below 20 GeV and steeper above, and results from
our analysis of Galactic diffuse \gray emission.  The ``background''
spectra are slightly dependent on the halo size.  Since all secondary
particles are produced in the Galactic plane, increasing the halo size
results only in a small decrease of the flux at high energies due to
larger energy losses.  The propagation parameters for these models are
given in Strong \& Moskalenko (1998) and Strong, Moskalenko, \&
Reimer (1999), and the formalism for calculation of
secondary positrons is described in Moskalenko \& Strong (1998).

\indent
We do not intend to make sophisticated calculations of positron spectra
resulting from numerous decay chains such as best done by, e.g., Baltz
\& Edsj\"o (1998) for many WIMP candidates.  Instead, for illustration
purposes, we simplify our analysis by treating the annihilation to
$W^\pm$ and $Z^0$-pairs.  For $m_\chi < m_W$ we consider only the
direct annihilation to $e^{+}e^{-}$ pairs.  In the first case we use
the cross sections for a pure Higgsino (Kamionkowski \& Turner 1991),
in the latter case we take $B\cdot \sigv =3\times 10^{-28}$ cm$^3$
s$^{-1}$ and monoenergetic positrons.  These parameters can be
considered as optimistic, but possible.  To maximize the signal we
further choose the Galactic halo size as $10$ kpc.
\end{figwindow}

Fig.~\ref{fig2} shows our predictions for the two CR positron
``background'' models together with HEAT data (Barwick et al.\ 1998) and
positrons from neutralino annihilation. It is seen that the predicted
signal/background ratio has a maximum near $m_\chi \sim m_W$, while
even in the ``conventional'' model the background is nearly equal to
the signal at its maximum.  It is however interesting to note that our
calculations in this model show some excess in low energy ($\leq 10$ GeV)
positrons where the measurements are rather precise but the solar
modulation is also essential. If this excess testifies to a corresponding
excess in  interstellar space and if the positron background
correspond to our ``conventional'' calculations, it could be a hint for
the presence of the dark matter (Baltz
\& Edsj\"o 1998, Coutu et al.\ 1999).  Our HEMN model fits
the HEAT data better (no excess) and thus provides more background
positrons.
(This shows that in principle a good fit to positron data, which is
consistent also with other measurements such as \grays and antiprotons
is possible without any additional positron source.)
Under such circumstances a significant detection of a weak
signal would require favourable conditions and precise
measurements. Though this conclusion qualitatively agrees with that
of Baltz and Edsj\"o (1998) and several earlier papers, it is
based on a more realistic model of particle propagation and thus reduces
the scope for future speculations.

\section{Conclusions:}
Our propagation model has been used to study several areas of high
energy astrophysics.  We use this model for the calculation of positron
propagation in different models of the dark matter halo distribution.
We have shown that the Green's functions are not very sensitive to the
dark matter distribution for the same local dark matter energy
density.  This is a natural consequence of the large positron energy
losses.  We compare our predictions with the computed CR positron
``background'' for two models of the CR nucleon spectrum.  A correct
interpretation of positron measurements requires reliable background
calculations and thus emphasizes the necessity for further developments
in modelling production and propagation of CR species in the Galaxy.


\vspace{1ex}
\begin{center}
{\Large\bf References}
\end{center}
Adriani, O., et al., 1997, Proc.\ 25th ICRC (Durban) 5, 49\\
Baltz, E.A., \& Edsj\"o, J.  1998, Phys.\ Rev.\ D 59, 023511\\
Barbiellini, G., et al.  1997, Proc.\ 25th ICRC (Durban) 5, 1\\
Barwick, S.W., et al.  1998, \apj 498, 779\\
Coutu, S., et al.  1999, Astropart.\ Phys., in press (\astroph{9902162})\\
Ellis, J. 1998, Proc.\ Nobel Symp.\ (Sweden) (\astroph{9812211})\\
Jungman, G., Kamionkowski, M., \& Griest, K.  1996, Phys.\ Rep.\ 267, 195\\
Kamionkowski, M., \& Kinkhabwala, A., 1998, Phys.\ Rev.\ D 57, 3256\\
Kamionkowski, M., \& Turner, M.S.  1991, Phys.\ Rev.\ D 43, 1774\\
Moskalenko, I.V., \& Strong, A.W.  1998, \apj 493, 694\\
Moskalenko, I.V., \& Strong, A.W.  1999, Phys.\ Rev.\ D, in press\\
Strong, A.W., \& Moskalenko, I.V.  1998, \apj 509, 212\\
Strong, A.W., \& Moskalenko, I.V.  1999, Proc.\ Workshop
   ``LiBeB, Cosmic Rays and Gamma-Ray Line Astronomy'', eds. R.Ramaty et
   al., ASP Conf.\ Ser.\ 171, 154\\
Strong, A.W., Moskalenko, I.V., \& Reimer, O.  1999, submitted
  (\astroph{9811296})\\
Trimble, V.  1989, Ann.\ Rev.\ Astron.\ Astrophys.\ 25, 425\\

\end{document}